\documentclass[11pt]{article}
\usepackage{moriond,epsfig}

\def\Journal#1#2#3#4{{#1} {\bf #2}, #3 (#4)}

\def\PLB{{\em Phys. Lett.}  B}

\def\PRD{{\em Phys. Rev.} D}

\def\be{\begin{equation}}
\def\ee{\end{equation}}
\def\bea{\begin{eqnarray}}
\def\eea{\end{eqnarray}}

\begin{document}
\vspace*{4cm}
\title{Inflation After WMAP}

\author{ William H. Kinney }

\address{
Dept. of Physics
University at Buffalo, SUNY
Buffalo, NY 14260-1500
}
 
\maketitle\abstracts{
I examine the status of inflationary cosmology in light of the first-year data from the WMAP satellite, focusing on the simplest models of inflation: those driven by a single scalar field. The WMAP observation of the Cosmic Microwave Background is the first unified, self-consistent data set capable of putting meaningful constraints on the inflationary parameter space. WMAP provides significant support for the inflationary paradigm in general, and single-field slow-roll inflation models provide a good fit to existing observational constraints. 
}

\section{Introduction}

Inflation\cite{inflationreviews} has emerged as the most successful model for understanding the physics of the very early universe. Inflation in its most general form consists of a period of accelerating expansion, during which the universe is driven toward flatness and homogeneity. In addition, inflation provides a mechanism for generating the initial perturbations which led to structure formation in the universe. The key ingredient of this cosmological acceleration is negative pressure, or a fluid with a vacuum-like equation of state $p \sim - \rho$. In order for inflation to end and the universe to transition to the radiation-dominated expansion necessary for primordial nucleosynthesis, this vacuum-like energy must be dynamic, and therefore described by one or more order parameters with quantum numbers corresponding to vacuum, {\em i.e.} scalar fields. In the absence of a compelling model for inflation, it is useful to consider the simplest models, those described by a single scalar order parameter $\phi$, with potential $V(\phi)$ and energy density and pressure for a homogeneous mode of
\begin{equation}
\rho = {1 \over 2} \dot\phi^2 + V\left(\phi\right),\ 
p = {1 \over 2} \dot\phi^2 - V\left(\phi\right).
\end{equation}
The negative pressure required for inflationary expansion is achieved if the field is slowly rolling, $\dot\phi^2 \ll V(\phi)$, so that the potential dominates. Different inflationary models can then be constructed by specifying different choices for the potential $V(\phi)$. In turn, different choices for $V(\phi)$ predict different spectra for primordial fluctuations in the universe, and precision observations can shed light on the physics relevant during the inflationary epoch. In this paper, I discuss the status of inflation in light of the WMAP observation of the Cosmic Microwave Background (CMB). The paper is arranged as follows: in Section \ref{seczooplot}, I discuss the observational predictions of inflation. In Section \ref{secWMAPconstraints} I discuss the constraints on the inflationary parameter space from the WMAP CMB anisotropy measurement. In Section \ref{secphi4} I discuss constraints on a particular potential, $V(\phi) = \lambda \phi^4$. Section \ref{secconclusions} contains a brief summary and conclusions.

\section{The zoo plot}
\label{seczooplot}

During inflation, quantum fluctuations on small scales
are quickly redshifted to scales much larger than the horizon size, where they
are ``frozen'' as perturbations in the background
metric. The metric perturbations created during inflation are of two types, both of
which contribute to CMB anisotropy:
scalar, or {\it curvature} perturbations, which couple to the stress-energy of
matter in the universe and form the ``seeds'' for structure formation, and
tensor, or gravitational wave perturbations, which do not couple to matter. Scalar
fluctuations can also be interpreted as fluctuations in the density of the matter in the universe, and can be
quantitatively characterized by perturbations $P_{\cal R}$ in the intrinsic
curvature scalar
\begin{equation}
P_{\cal R}^{1/2}\left(k\right) = {1 \over \sqrt{\pi}} {H \over M_{Pl}
\sqrt{\epsilon}}\Biggr|_{k^{-1} = d_H}.
\end{equation}
The fluctuation power is in general a function of wavenumber $k$, and is
evaluated when a given mode crosses outside the horizon during inflation,
$k^{-1} = d_H$. The {\em slow roll parameter} $\epsilon$ is defined by the 
variation in the Hubble parameter with field value, and for a slowly rolling field
($\dot\phi^2 \ll V(\phi)$) 
is given approximately in terms of the first derivative of the potential by:
\begin{equation}
\epsilon = {m_{\rm Pl}^2 \over 4 \pi} \left({H'\left(\phi\right) \over H\left(\phi\right)}\right)^2 \simeq {m_{\rm Pl}^2 \over 16 \pi} \left({V'\left(\phi\right) \over V\left(\phi\right)}\right)^2.
\end{equation}
This parameter governs the equation of state of the scalar field as
\begin{equation}
p = \rho \left({2 \over 3} \epsilon - 1\right),
\end{equation}
so that accelerating expansion occurs for $\epsilon < 1$, or $p < -\rho/3$.
The {\em spectral index} $n$ is defined by assuming an
approximately power-law form for $P_{\cal R}$ with
\begin{equation}
n - 1 \equiv {d\ln\left(P_{\cal R}\right) \over d\ln\left(k\right)} \simeq - 4 \epsilon + 2 \eta,
\end{equation}
so that a scale-invariant spectrum, in which modes have constant amplitude at
horizon crossing, is characterized by $n = 1$. Here $\eta$ is the second slow roll parameter,
\begin{equation}
\eta\left(\phi\right) \equiv {M_{Pl}^2 \over 4 \pi} \left({H''\left(\phi\right)
\over H\left(\phi\right)}\right) \simeq {M_{Pl}^2 \over 8 \pi}
\left[{V''\left(\phi\right) \over V\left(\phi\right)} - {1 \over 2}
\left({V'\left(\phi\right) \over V\left(\phi\right)}\right)^2\right].
\end{equation}
Variation of the spectral index $dn/d\ln(k)$ with scale is second order in slow-roll,
{\em i.e.} of order $\epsilon^2$.
Similarly, the power spectrum of tensor fluctuation modes is given
by
\begin{equation}
P_{T}^{1/2}\left(k\right) = {4 \over \sqrt{\pi}} {H \over M_{Pl}}\Biggr|_{k^{-1} = d_H}.
\end{equation}
The ratio of tensor to scalar modes is then
\begin{equation}
{P_{T} \over P_{\cal R}} =  16 \epsilon,
\end{equation}
so that tensor modes are negligible for $\epsilon \ll 1$. Tensor and scalar modes both contribute to CMB temperature anisotropy. The tensor spectral index is 
\begin{equation}
n_{T} \equiv {d \ln\left(P_{T}\right) \over d\ln\left(k\right)} = - 2 \epsilon.
\end{equation}
Note that $n_{T}$ is {\it not} an independent parameter, but is proportional to the tensor/scalar ratio,
known as the {\it consistency relation} for single-field inflation. Many papers in
the recent literature define the tensor/scalar ratio directly in terms of the primordial power spectra, $r \equiv \left(P_{T} / P_{\cal R}\right)$, which has the
advantage of being independent of other cosmological parameters. Another convention defines the tensor/scalar ratio in terms of the relative contribution of fluctuations to the CMB power at a particular scale, usually the quadrupole,
\begin{equation}
r \equiv {C^{\rm tensor}_2 \over C^{\rm scalar}_2} \simeq 10 \epsilon,
\end{equation} 
where the relationship to the slow roll parameter $\epsilon$ depends on the matter/dark energy content of the universe. The above figure $r \simeq 10 \epsilon$ is valid for $\Omega_{\rm M} \simeq 0.3$ and $\Omega_\Lambda \simeq 0.7$. I use the latter convention in this paper. 

\begin{figure}
\centerline{\psfig{figure=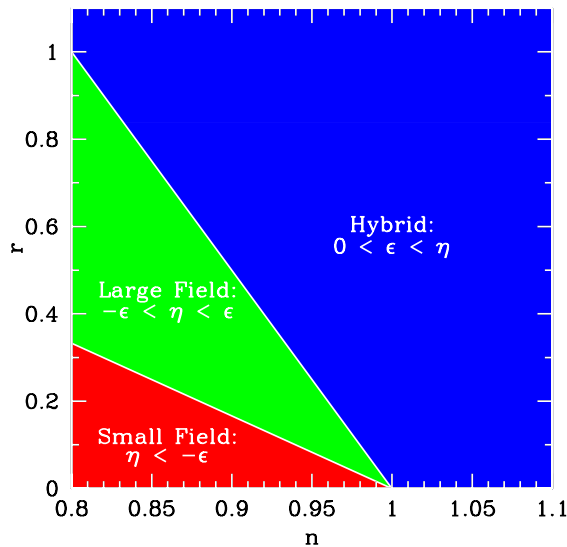,height=3.0in}}
\caption{The ``zoo plot'' of inflationary models in the $n-r$ plane.
\label{fig:zooplot}}
\end{figure}

A given inflation model can therefore be described to lowest order in slow roll by three independent parameters, $P_{\cal R}$, $P_{T}$, and $n$. To next order in slow roll, we add the running of the spectral index, $dn/d\ln(k)$. The overall normalization is typically fixed by a free parameter in the inflationary potential, so that the parameters relevant for distinguishing among inflationary parameters are the tensor/scalar ratio $r$, the scalar spectral index $n$, and the running $dn/d\ln(k)$. Different choices for the inflationary potential result in different predictions for the parameters, and therefore constraints on the power spectrum from the CMB can be used to rule out inflationary models.\cite{Dodelson:1997hr,Kinney:1998md} It is useful to divide inflation models into three broad classes defined by the relationship between the slow roll parameters $\epsilon$ and $\eta$: {\em small field} models, with $\eta < - \epsilon$, {\em large-field} models, with $-\epsilon < \eta < \epsilon$, and {\em hybrid} models, with $\eta > \epsilon$. Typical potentials for small-field models are of the form $V(\phi) = \Lambda^4 \left[1 - (\phi / \mu)^p\right]$, for example models based on spontaneous symmetry breaking phase transitions where the field rolls away from an unstable equilibrium at $\phi = 0$. Typical large-field potentials are of the form $V(\phi) = \Lambda^4 (\phi / \mu)^p$, where the field during inflation has value $\phi > M_{Pl}$ and rolls toward the origin. Typical hybrid-type models have potentials of the form $V(\phi) = \Lambda^4 \left[1 + (\phi / \mu)^p\right]$, and require an auxiliary field to end inflation.\cite{Linde:1993cn} This ``zoology'' of models is useful because the three classes occupy different regions of the plane of observable parameters $n$ and $r$, shown in Fig. \ref{fig:zooplot}. 

A complementary procedure to this classification scheme is to generate large numbers of particular inflation models by Monte Carlo evaluation of an infinite hierarchy of {\em flow equations} which completely specify the inflationary dynamics\cite{flowpapers}
\begin{eqnarray}
{d \epsilon \over d N} &=& \epsilon \left(\sigma + 2 \epsilon\right),\cr
{d \sigma \over d N} &=& - 5 \epsilon \sigma - 12 \epsilon^2 + 2
 \left({}^2\lambda_{\rm H}\right),\cr
{d \left({}^\ell\lambda_{\rm H}\right) \over d N} &=& \left[{1 \over 2}
 \left(\ell - 1\right) \sigma + \left(\ell - 2\right) \epsilon\right]
 \left({}^\ell\lambda_{\rm H}\right) + {}^{\ell+1}\lambda_{\rm
 H},\label{eqfullflowequations}
\end{eqnarray}
where $N$ is the number of e-folds of inflation, $\sigma \equiv 4 \epsilon - 2 \eta$, and the higher-order slow roll parameters are defined by\cite{liddle94}
\begin{equation}
{}^\ell\lambda_{\rm H} \equiv \left({m_{\rm Pl}^2 \over 4 \pi}\right)^\ell
 {\left(H'\right)^{\ell-1} \over H^\ell} {d^{(\ell+1)} H \over d\phi^{(\ell +
 1)}}.\label{eqdefoflambda}
\end{equation}
Since the dynamics are governed by a set of first-order differential equations, the
cosmological evolution is entirely specified by choosing values for the
slow roll parameters $\epsilon, \sigma, {}^2\lambda_{\rm H},\ldots$. Choosing
such a point in the parameter space completely specifies the inflationary
model, including the scalar field potential, which can be reconstructed
for each choice, so-called ``Monte Carlo reconstruction''. \cite{montecarlorecon} Thus it is possible to generate an ensemble of potentials consistent with a particular observational constraint. (See the referenced papers for a detailed explanation of the technique.)

\section{Constraints from WMAP}
\label{secWMAPconstraints}

\begin{figure}
\centerline{\psfig{figure=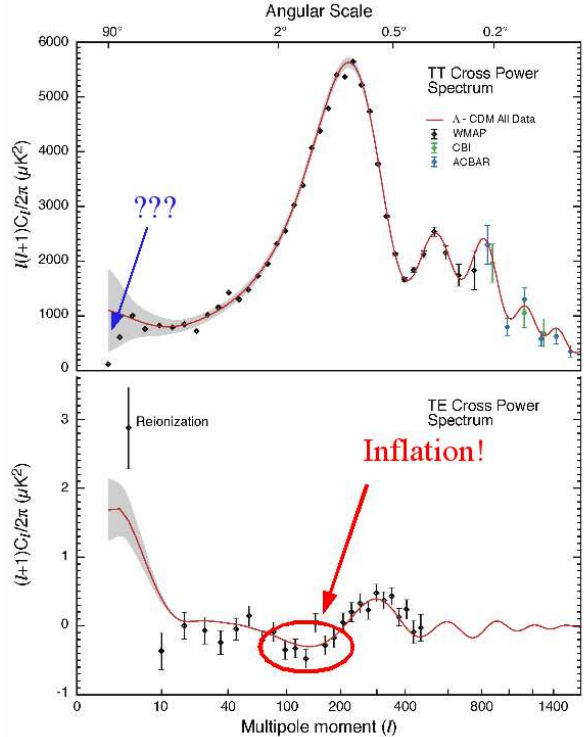,height=4.0in}}
\caption{The temperature autocorrelation and temperature/polarization cross-correlation multipole spectra from WMAP. The measured anticorrelation in the cross-correlation spectrum is around $\ell = 100$ is a strong indication of inflation. (Figure courtesy of the NASA/WMAP science team.)
\label{fig:WMAPspectrum}}
\end{figure}

The spectacular WMAP observation has delivered on the promise of precision cosmology. Figure \ref{fig:WMAPspectrum} shows the famous angular power spectrum for the CMB temperature and polarization from WMAP. 
Two features are marked on Fig. \ref{fig:WMAPspectrum}. The first is the well-known deficit in the power spectrum at large angles (low $\ell$). This puzzling feature has yet to be definitively explained and is problematic for slow-roll inflation models, which tend to predict a nearly power-law spectrum of primordial fluctuations. A second feature, less commented upon in the literature, is the ``dip'' in the temperature/polarization cross-correlation spectrum between $\ell = 100$ and $\ell = 200$. Unlike the low quadrupole, this anticorrelation in the temperature and polarization signals is exactly what is predicted for a background of adiabatic fluctuations produced by inflation. Adiabatic modes are readily distinguished topological defects, which produce a positive correlation on similar scales.\cite{Peiris:2003ff} Furthermore, the scales corresponding to $\ell = 100$ are larger than the horizon size at the time of recombination. Therefore this signal in the temperature/polarization cross-correlation spectrum is a ``smoking gun'' for the presence of correlations in the primordial fluctuations in {\em causally disconnected} regions, a distinct signature of inflation. \cite{Spergel:1997vq}  Like the flatness of the universe, this is an essentially model-independent prediction of inflation, and in this sense WMAP has provided substantial and nontrivial support for the inflationary paradigm.

\begin{figure}
\centerline{\psfig{figure=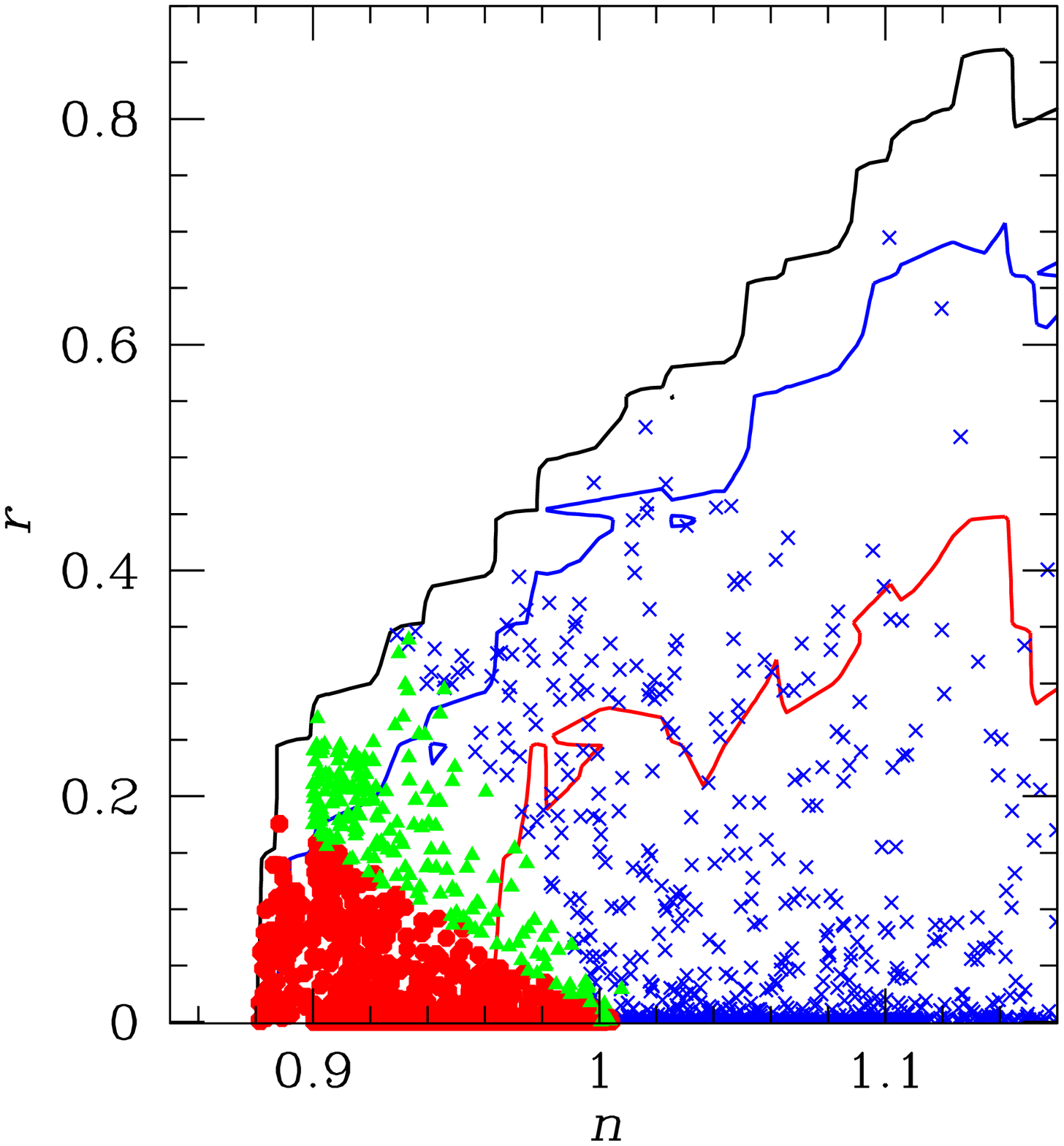,height=3.0in}
\psfig{figure=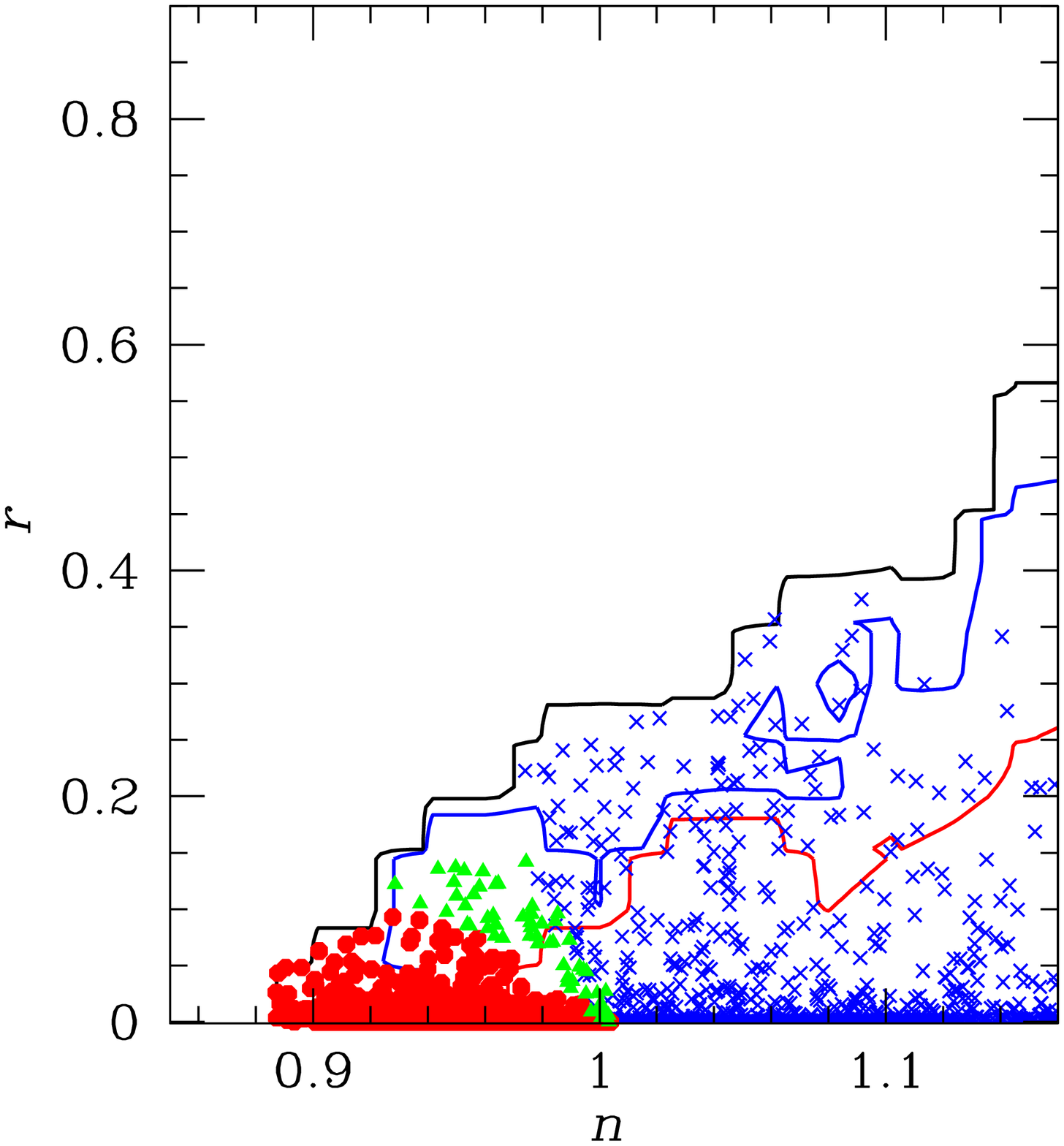,height=3.0in}}
\caption{
CMB constraints on the $n-r$ plane from WMAP (left) and WMAP+7 (right). The points are models generated by Monte Carlo, and are coded by type: small field (red, circles), large-field (green, triangles), and hybrid (blue, crosses). Only models consistent with the data to within a $3\sigma$ likelihood are shown. The contours represent $1\sigma$, $2\sigma$, and $3\sigma$ likelihoods. These contours are {\em projections} of the error bars in the full 3-D parameter space $r$, $n$, $dn/d\ln(k)$: the third parameter is not marginalized over. 
\label{fig:anvsr}}
\end{figure}

\begin{figure}
\centerline{\psfig{figure=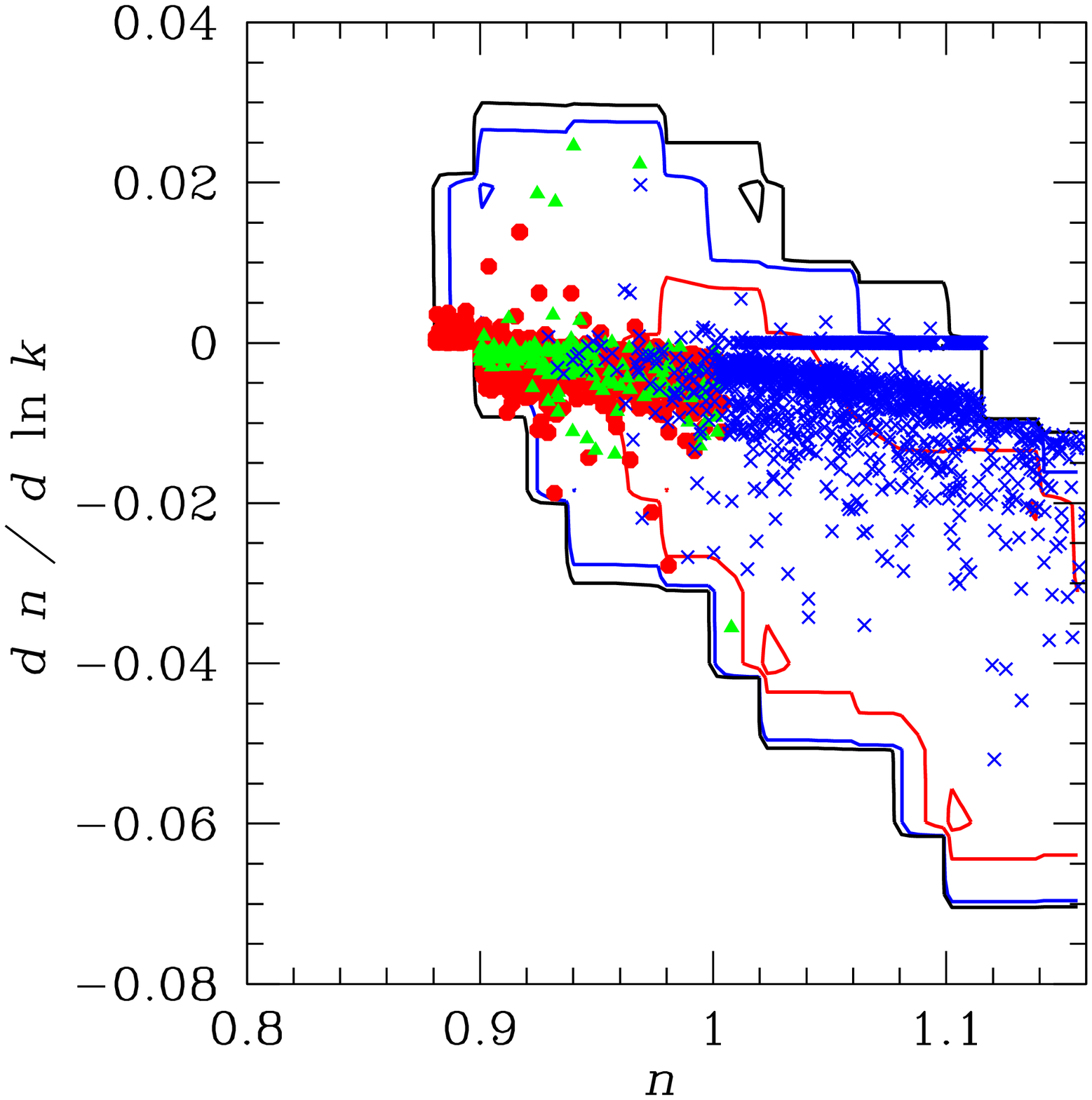,height=3.0in}
\psfig{figure=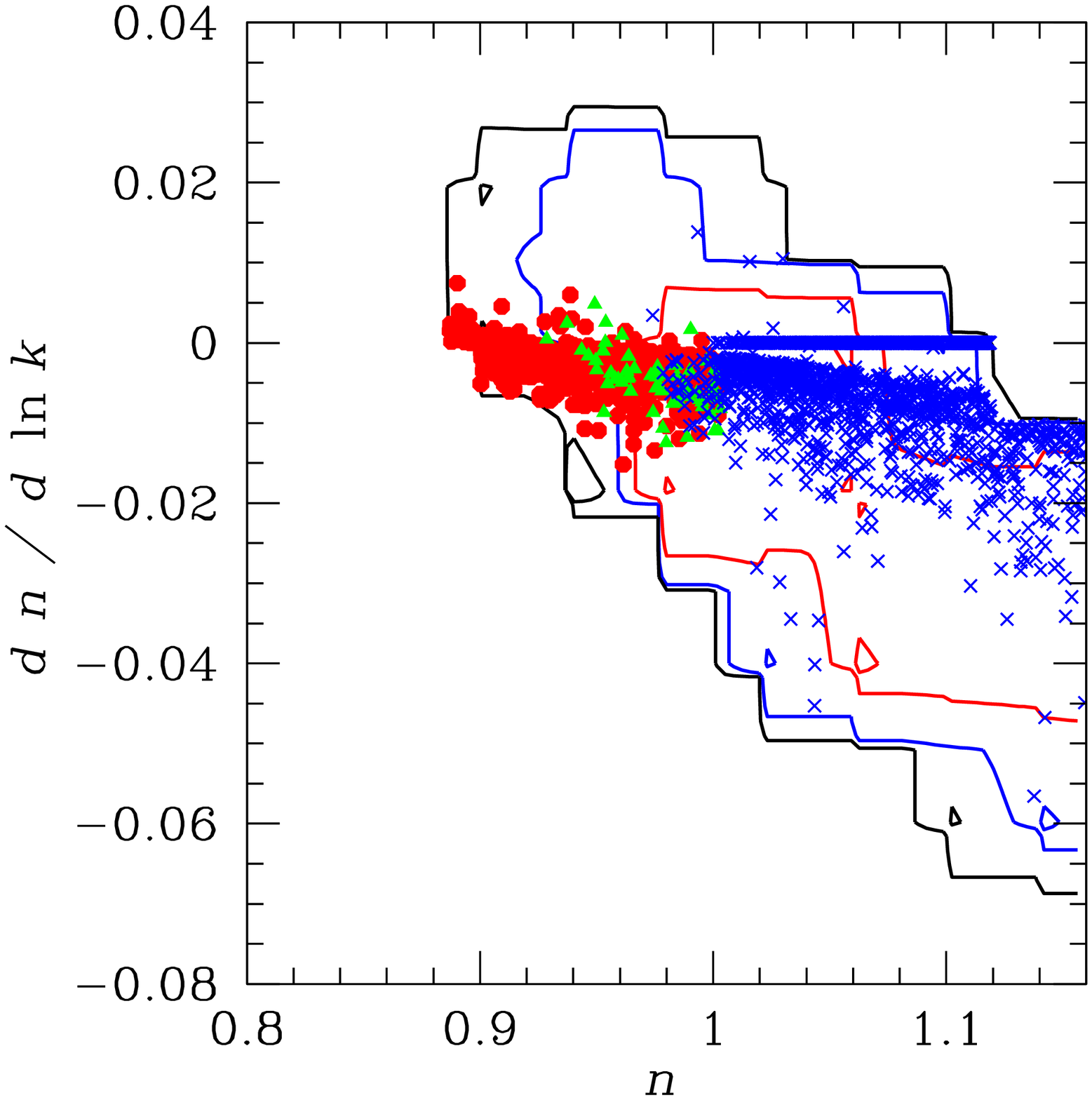,height=3.0in}}
\caption{
CMB constraints on the $n-dn/d\ln(k)$ plane from WMAP (left) and WMAP+7 (right). The points are models generated by Monte Carlo, and are coded by type: small field (red, circles), large-field (green, triangles), and hybrid (blue, crosses). The contours represent $1\sigma$, $2\sigma$, and $3\sigma$ likelihoods. 
\label{fig:anvsdn}}
\end{figure}

The WMAP data is additionally significant for inflation because it is the first unified, self-consistent data set capable of putting meaningful constraints on the inflationary parameter space.\footnote{Previously, balloon or ground-based data had to be combined with the large angular scale data from COBE to constrain the power spectrum\cite{Kinney:2000nc,Hannestad:2001nu}. This left the parameters subject to uncertainties due to calibration errors and possibly other systematics.} While it is useful to augment the WMAP data set by combining it with other data such as large-scale structure surveys, it is also desirable to understand what this data set alone tells us about inflation. Figures \ref{fig:anvsr} and \ref{fig:anvsdn} show constraints on the space of inflation paramters $r$, $n$, and $dn/d\ln(k)$ for WMAP alone and for WMAP combined with seven other CMB anisotropy measurements: Boomerang, MAXIMA-1, DASI, CBI, ACBAR, VSA, and ARCHEOPS.\cite{Kinney:2003uw} (I will refer to this second data set at ``WMAP+7''.) The data are fit to seven parameters: the dark matter density $\Omega_{\rm M} h^2$, the baryon density $\Omega_{\rm b} h^2$, the dark energy density $\Omega_{\Lambda} h^2$, the reionization optical depth $\tau$, and the three inflationary parameters $r$, $n$, and $dn/d\ln(k)$. The dark energy is assumed to be a cosmological constant, with equation of state $p = - \rho$. Four priors are assumed on the parameters space: (1) flat universe $\Omega_{\rm tot} = 1$, (2) HST key project value for the Hubble constant, $h = 0.72 \pm 0.15$, (3) reionization optical depth $\tau < 0.3$, and (4) the inflationary consistency condition for the tensor spectral index $n_{\rm tensor} = - r / 5$. Figure \ref{fig:anvsr} shows constraints the $n-r$ plane for WMAP and WMAP+7, along with models generated by Monte Carlo as described in Sec. \ref{seczooplot}.

The first clear conclusion to be drawn from the constraints is that the single-field, slow-roll inflation scenario is not strongly challenged by the existing CMB data taken alone. In particular, the ``vanilla'' scale-invariant ($n = 1$), tensor-free ($r = 0$) power spectrum with no running is consistent with both the WMAP and WMAP+7 data sets. The best-fit model prefers a negative running, $dn/d\ln(k) < 0$, consistent with the Peiris, {\it et al.} result, \cite{Peiris:2003ff} but the preference is not statistically significant. (The Peiris, {\it et al.} result indicating a roughly $2\sigma$ indication of negative running was arrived at by combining WMAP with 2dFGRS and Lyman-$\alpha$ forest data to give a longer ``lever arm'' on the power spectrum.\cite{Peiris:2003ff}) The models plotted in the $n-r$ plane (Fig. \ref{fig:anvsr}) show no particular preference for small-field, large-field or hybrid: all three classes are consistent with the data within the $1\sigma$ likelihood contour. However, considering the running of the spectral index (Fig. \ref{fig:anvsdn}) reveals a weak preference for hybrid models, with more fully populate the best-fit region with $dn/d\ln(k) < 0$.

\section{Is $\lambda \phi^4$ dead?}
\label{secphi4}

As a concrete example of the discriminatory power of current data, consider a particular potential, $V\left(\phi\right) = \lambda \phi^4$. This is a useful example because, in addition to being one of the simplest possible inflationary potentials, the predicted power spectrum lies very close to the border between the regions allowed and disallowed by observation. Therefore the influence of priors on the result is especially important, as is the choice of data sets used to constrain the model. 

One important factor when calculating the prediction of the model is the choice of the number of e-folds $N$.
For $V \propto \phi^4$, the slow roll parameters depend on $N$ as
$\epsilon = 2 \eta = 1 /  \left(N + 1\right)$.
Thus the predictions for the observable parameters $r$, $n$, and $dn/d\ln(k)$ depend on $N$, with
\begin{equation}
r \simeq {10 \over N + 1},\ 
n \simeq 1 - {3 \over N + 1},\ 
{dn \over d \ln(k)} \simeq - {3 \over N (N + 1)}.
\end{equation}
This is to lowest order in slow roll, where I have used the normalization convention $r = 10 \epsilon$ \cite{Kinney:2003uw}. $N$ is the number of e-folds before the end of inflation which corresponds to the horizon-crossing time of a mode with a wavelength equal to the pivot scale today, and it cannot be fixed {\it a priori}. The relationship between $N$ and perturbation wavelength depends on details such as reheat temperature, which cannot be fixed without a more detailed model. 
Relative to the WMAP+7 data set, $N = 40$ is ruled out to high significance. $N < 66$ is  ruled out at $3 \sigma$ relative to the error contour in the full three-dimensional parameter space. Constraints from WMAP alone are much less stringent, with $N < 40$ ruled out to $3 \sigma$, and $N \geq 40$ consistent with the constraint from WMAP alone. A number of groups have investigated similar constraints, with different choices of data set and priors:

\begin{itemize}
\item{Peiris, {\it et al.}:\cite{Peiris:2003ff} $\lambda \phi^4$ ruled out for $N = 50$, using WMAP + 2dF Galaxy Redshift Survey (2dFGRS)\cite{Percival:2001hw} + Lyman-$\alpha$ forest.}
\item{Barger, {\it et al.}:\cite{Barger:2003ym} $\lambda \phi^4$ OK for $N > 45$, using WMAP only and assuming $dn /d \ln(k) = 0$.}
\item{Leach and Liddle:\cite{Leach:2003us} $\lambda \phi^4$ OK for $N \geq 60$, using WMAP + 2dFGRS.}
\item{WHK, {\it et al.}:\cite{Kinney:2003uw} $\lambda \phi^4$ OK for $N > 40$ (WMAP only), and ruled out for $N < 66$ (WMAP + 7).}
\item{Tegmark, {\it et al.}:\cite{Tegmark:2003ud} $\lambda \phi^4$ marginal for WMAP + Sloan Digital Sky Survey}. 
\end{itemize}

It is encouraging that a number of independent analyses, using different data sets and priors, arrive at comparable conclusions about constraints on the space of inflation parameters. It is also clear that the $\lambda \phi^4$ model is under significant pressure from the data. It cannot, however, be said to be definitively ruled out by current constraints.

\section{Conclusions}
\label{secconclusions}

Inflationary cosmology, in light of recent data, is in very good shape. The WMAP temperature/polarization cross-correlation spectrum provides clear evidence for the presence of adiabatic density perturbations correlated on scales larger than the horizon size at the time of recombination, strong support for the inflationary paradigm. WMAP constraints on the primordial power spectrum are consistent with the predictions of the simplest single-field inflation models, including the most generic inflationary prediction of a scale-invariant spectrum with no detectable contribution from tensor fluctuations. This is the case for the WMAP data taken alone or in combination with other CMB observations. The best-fit value of a blue ($n > 1$) scalar power spectrum with a negative running of the scalar spectral index is consistent with results using WMAP combined with constraints from large-scale structure.\cite{Peiris:2003ff,Leach:2003us} Particular choices of inflationary potential, for example $V(\phi) = \lambda \phi^4$, are strongly challenged by the data: observational cosmology has reached the point where it is possible to rule out models of the very earliest moments of the universe, a significant achievement. Some anomalies remain, most notably the unusually low signal near the quadrupole, which would be difficult for the simplest inflationary models to explain. Future data, such as large-scale structure data from the Sloan Digital Sky Survey, ground-based CMB polarization measurements, and the Planck satellite, will significantly improve the constraints on the inflationary parameter space and meaningfully test the inflationary paradigm as a whole. The era of precision cosmology is here. 

\section*{Acknowledgments}

Much of the work summarized here was done in collaboration with Alessandro Melchiorri, Edward W. Kolb, and Antonio Riotto\cite{Kinney:2003uw}. I also thank Richard Battye for useful conversations.


\section*{References}
\begin{thebibliography}{99}


\bibitem{inflationreviews}
For review articles on inflation with more complete reference lists, see:
D.~H.~Lyth and A.~Riotto, 
\Journal{\em Phys. Rept.}{314}{1}{1999} [arXiv:hep-ph/9807278];\\
W.~H.~Kinney,
arXiv:astro-ph/0301448.

\bibitem{Dodelson:1997hr}
S.~Dodelson, W.~H.~Kinney and E.~W.~Kolb,
\Journal{PRD}{56}{3207}{1997}
[arXiv:astro-ph/9702166].

\bibitem{Kinney:1998md}
W.~H.~Kinney,
\Journal{PRD}{58}{123506}{1998} 
[arXiv:astro-ph/9806259].

\bibitem{Linde:1993cn}
A.~D.~Linde,
\Journal{PRD}{49}{748}{1994}
[arXiv:astro-ph/9307002].

\bibitem{flowpapers} M.~B.~Hoffman and M.~S.~Turner, 
\Journal{PRD}{64}{023506}{2001} [arXiv:astro-ph/0006321];\\
 D.~J.~Schwarz, C.~A.~Terrero-Escalante, and A.~.A.~Garcia,
\Journal{PLB}{517}{243}{2001} [arXiv: astro-ph/0106020];\\
W.~H.~Kinney, 
\Journal{PRD}{66}{083508}{2002} [arXiv:astro-ph/0206032].

\bibitem{liddle94}
A.~R.~Liddle, P.~Parsons, and J.~D.~Barrow,
\Journal{PRD}{50}{7222}{1994}
[arXiv:astro-ph/9408015].

\bibitem{montecarlorecon}
W.~H.~Kinney and R.~Easther, 
\Journal{PRD}{67}{043511}{2003}
[arXiv:astro-ph/0210345].

\bibitem{Peiris:2003ff}
H.~V.~Peiris {\it et al.},
\Journal{\em Astrophys.\ J.\ Suppl.}{148}{213}{2003}
[arXiv:astro-ph/0302225].

\bibitem{Spergel:1997vq}
D.~N.~Spergel and M.~Zaldarriaga,
\Journal{PRL}{79}{2180}{1997}
[arXiv:astro-ph/9705182].

\bibitem{Kinney:2000nc}
W.~H.~Kinney, A.~Melchiorri and A.~Riotto,
\Journal{PRD}{63}{023505}{2001} 
[arXiv:astro-ph/0007375].

\bibitem{Hannestad:2001nu}
S.~Hannestad, S.~H.~Hansen, F.~L.~Villante and A.~J.~S.~Hamilton,
\Journal{\em Astropart.\ Phys.}{17}{375}{2002}
[arXiv:astro-ph/0103047].

\bibitem{Kinney:2003uw}
W.~H.~Kinney, E.~W.~Kolb, A.~Melchiorri and A.~Riotto,
\Journal{\PRD}{69}{103516}{2004}
[arXiv:hep-ph/0305130].

\bibitem{Percival:2001hw}
W.~J.~Percival {\it et al.},
\Journal{Mon.\ Not.\ Roy.\ Astron.\ Soc.}{327}{1297}{2001}.
[arXiv:astro-ph/0105252].

\bibitem{Barger:2003ym}
V.~Barger, H.~S.~Lee and D.~Marfatia,
\Journal{\PLB}{565}{33}{2003}
[arXiv:hep-ph/0302150].


\bibitem{Leach:2003us}
S.~M.~Leach and A.~R.~Liddle,
\Journal{\PRD}{68}{123508}{2003}
[arXiv:astro-ph/0306305].

\bibitem{Tegmark:2003ud}
M.~Tegmark {\it et al.}  [SDSS Collaboration],
\Journal{\PRD}{69}{103501}{2004}
[arXiv:astro-ph/0310723].

\end{thebibliography}
\end{document}